\documentclass[prl,amsmath,amssymb,dvips]{revtex4}

\usepackage{graphics,epsfig,amssymb}
\usepackage{hyperref}

\def\beq{\begin{equation}} 
\def\eeq{\end{equation}} 
\begin{document}

\title{Exact $T=0$ Eigenstates of the Isovector Pairing Hamiltonian}

\author{M. Sambataro$^a$ and N. Sandulescu$^b$}
\affiliation{$^a$Istituto Nazionale di Fisica Nucleare - Sezione di Catania,
Via S. Sofia 64, I-95123 Catania, Italy \\
$^b$National Institute of Physics and Nuclear Engineering, P.O. Box MG-6, 
Magurele, Bucharest, Romania}

\begin{abstract}
We derive
the exact $T=0$ seniority-zero eigenstates of the isovector pairing Hamiltonian
for an even number of protons and neutrons. Nucleons are supposed to be distributed over a set of non-degenerate levels and to interact through a pairing force with constant strength. We show that these eigenstates (and among them, in particular, the ground state) are linear superpositions of products of $T=1$ collective pairs arranged into $T=0$ quartets. 
This grouping of protons and neutrons first into $T=1$ collective pairs and 
then into $T=0$ quartets represents the distinctive feature of these eigenstates.
This work highlights, for the first time on the grounds of the analytic expression of its eigenstates, the key role played by the isovector pairing force in the phenomenon of nuclear quarteting. 
\end{abstract}

\maketitle

\section{Introduction}
Exact analytic treatments of model Hamiltonians can provide very useful insights into important 
phenomena occurring in quantum many-body systems. A model Hamiltonian which has received much attention over the years in the context of nuclear structure is the isovector pairing Hamiltonian \cite{frau}. This Hamiltonian describes a two-component system consisting of protons $(p)$ and neutrons $(n)$ interacting through a pairing force which acts on an equal footing on $pp$, $nn$ and $pn$ pairs.
Particularly in light nuclei, owing to the fact that protons and neutrons share the same orbitals and due to the charge independence of the nuclear force, this Hamiltonian
represents an essential component of the effective nuclear interaction. 

An exact analytic treatment of the isovector pairing Hamiltonian for a set of non-degenerate levels 
has proved to be problematic.  The oldest approach dates back to the 60's and is due to Richardson \cite{richa1}. This formalism was further elaborated a few years later by the same author in collaboration with Chen \cite{chen}. More than three decades later, however,
Pan and Draayer \cite{feng} showed that the approach
of Refs. \cite{richa1,chen}  was valid only for systems with at most two pairs of nucleons. They also proposed a different approach but explicitly develop their formalism for up to three pairs only. In the same year, Links et al. \cite{links} provided an exact eigenspectrum of 
the isovector Hamiltonian by applying the quantum inverse scattering method without, however, explicitly discussing its eigenstates. Finally, in 2006, Dukelsky et al. \cite{duke} presented 
a complete exact treatment of a more general isovector pairing Hamiltonian 
including also isospin-breaking terms.

As compared with the exact Richardson solution of the like-particle pairing \cite{richa0,richa3,richa2}, simply formulated as a product of distinct collective pairs,
the existing solution of the isovector pairing Hamiltonian appears much less transparent. 
Indeed, in this case the eigenstates are expressed not only by products of collective pairs but also in terms of isospin and special raising operators \cite{duke}. Furthermore, besides the standard pair energies associated with the Richardson pairs (see below), these eigenstates  also depend on an additional set of spectral parameters that do not have a clear physical interpretation. These facts prevent a simple understanding of the type of correlations induced by the isovector pairing force. 

The aim of this article is to propose a new exact treatment of the isovector pairing Hamiltonian in which the eigenstates  have a transparent physical interpretation.
Our approach has been focused on the $T=0$ seniority-zero eigenstates of a system with an even number of protons and neutrons. It will be shown that these eigenstates (and among them, in particular, the ground state) are linear superpositions of products of $T=1$ collective pairs arranged into $T=0$ quartets. 
The formation of such  $\alpha$-like four-body structures in the $T=0$ eigenstates of the isovector pairing Hamiltonian has not been pointed out in any of the previous exact treatments of this Hamiltonian. This result 
establishes, for the first time on the grounds of the analytic expression of its eigenstates, the key role played by the isovector pairing force in the emergence of nuclear quarteting, which is one of the oldest and still open issues in nuclear structure 
\cite{soloviev,flowers,ginocchio,faraggi,yamamura,catara,qm_prl}. 

The paper is structured as follows. In Section 2, we describe the formalism to construct the exact eigenstates. In Section 3, we provide some numerical applications. Finally, in Section 4, we give the conclusions. 

\section{The formalism}
The Hamiltonian under study reads as
\begin{equation}
H=\sum^\Omega_{i=1}\epsilon_i{\cal N}_i-g\sum^\Omega_{i,i'=1}\sum^{1}_{M_T=-1}P^{\dag}_{iM_T}P_{i'M_T},
\label{1}
\end{equation}
where
\begin{equation}
{\cal N}_i=\sum_{\sigma =\pm ,\tau =\pm\frac{1}{2}}a^\dag_{i\sigma\tau}a_{i\sigma \tau},~~~~
P^\dag_{iM_T}=[a^\dag_{i+}a^\dag_{i-}]^{T=1}_{M_T}.
\label{2}
\end{equation}
This Hamiltonian describes a system of protons and neutrons distributed over a set of $\Omega$ levels and interacting via an isovector pairing force with a level-independent strength $g$.
The operator $a^\dag_{i\sigma\tau}$ 
($a_{i\sigma\tau}$) creates (annihilates) a nucleon in the single-particle state characterized by the quantum numbers
$(i, \sigma ,\tau )$, where $i$ identifies one of the $\Omega $ levels of the model, 
$\sigma =\pm$ labels states which are conjugate with respect to time reversal and
$\tau =\pm\frac{1}{2}$ is the projection of the isospin of the nucleon. These operators obey standard fermion commutation relations.
The operator $P^\dag_{iM_T}$ $(P_{iM_T})$ creates (annihilates) a pair of nucleons in time-reversed states with total isospin $T=1$ and projection $M_T$. Depending on $M_T$, $P^\dag_{iM_T}$ creates a $pp$, a $nn$ or a $pn$ pair and the Hamiltonian $(\ref{1})$ is seen to act equally on these pairs.
Finally the operator ${\cal N}_i$ counts the number of nucleons on the level $i$, each level having an energy $\epsilon_i$. 

Owing to the presence of the $\sigma ,  \tau$ degrees of freedom, each level $i$ is fourfold degenerate being able to accommodate two protons and two neutrons in time-reversed states. We limit the Hilbert space of the model to states with total seniority-zero according to the notation of \cite{richa1}. For a $2N$-particle system the most geneneral seniority-zero space is spanned by the states
\begin{equation}
P^\dag_{i_1M_{T_1}}P^\dag_{i_2M_{T_2}}\cdots P^\dag_{i_NM_{T_N}}|0\rangle ,
\label{3}
\end{equation}
where $|0\rangle$ is the vacuum of the model. Since, as anticipated, we focus only on $T=0$ eigenstates, these states are subject to the condition 
$M_{T_1}+M_{T_2}+\dots +M_{T_N}=0$. The Hamiltonian (\ref{1}) does not mix states with different seniorities \cite{richa1}.

We begin by illustrating the formalism that we have adopted to construct the $T=0$ seniority-zero eigenstates of the Hamiltonian (\ref{1}) in the cases of
$2p-2n$ and $4p-4n$ systems. We will then discuss the generic case of a $Np-Nn$ system ($N$ even).

As a basic principle, following Richardson's suggestion \cite{richa1}, we assume the collective isovector pairs 
\begin{equation}
B^{\dag}_{\nu M_T}=\sum^{\Omega}_{k=1}\frac{1}{2\epsilon_k-E_{\nu}}P^{\dag}_{kM_T}
\label{4}
\end{equation}
as building blocks of the eigenstates of the Hamiltonian (\ref{1}).
These pairs are formally identical to those employed in the treatment of the like-particle pairing \cite{richa0,richa3,richa2}, differing only for the explicit  presence of the isospin degree of freedom. Their amplitudes depend on a parameter, $E_\nu$, that we shall name ``pair energy" as in the like-particle case.

For a $2p-2n$  system, according to the above principle, the only $T=0$ state that can be formed is
\begin{equation}
|\Psi^{(2)}\rangle =[B^\dagger_1B^\dagger_2]^0|0\rangle .
\label{11}
\end{equation}
This state is simply the product of two $T=1$ pairs (\ref{4}) coupled to $T=0$. In the following we will refer to such an $\alpha$-like structure as a $T=0$ quartet.
By making use of the commutation relations
\begin{equation}
[H,P^\dagger_{kM}]|0\rangle =(2\epsilon_kP^\dagger_{kM}-gP^\dagger_{M})|0\rangle 
\label{13}
\end{equation}
\begin{equation}
[[H,P^\dagger_{k_1M_1}],P^\dagger_{k_2M_2}]=
g\delta_{k_1,k_2}\sum_{M,M'}C(M_1,M_2,M,M')P^\dagger_{M}P^\dagger_{k_2M'},
\label{14}
\end{equation}
one can easily verify that
\begin{eqnarray}
H|\Psi^{(2)}\rangle&=&(E_1+E_2)|\Psi^{(2)}\rangle\nonumber\\
&&+\biggl\{1-\sum_i\frac{g}{2\epsilon_i-E_1}-\frac{g}{E_2-E_1}\biggr\}
[P^\dagger B^\dagger_2 ]^0|0\rangle\nonumber\\
&&+\biggl\{1-\sum_i\frac{g}{2\epsilon_i-E_2}-\frac{g}{E_1-E_2}\biggr\}
[P^\dagger B^\dagger_1 ]^0|0\rangle
\label{12}
\end{eqnarray}
where 
\begin{equation}
P^\dagger_M =\sum_kP^\dagger_{kM}.
\label{19}
\end{equation}
The matrix $C(M_1,M_2,M,M')$ in Eq. (\ref{14}) can be found in Ref. \cite{richa1}.
One deduces from the last expression that $|\Psi^{(2)}\rangle$ is an eigenstate of $H$ with eigenvalue $E^{(2)}=E_1+E_2$ if the pair energies $E_1$ and $E_2$ (the only parameters present in the definition (\ref{11})) are such that the two polynomials in curly brackets  are zero. This defines a system of two coupled non-linear equations in the two unknowns $E_1$ and $E_2$. In this case the present approach and Richardson's one \cite{richa1} coincide.

As a next case we consider the $4p-4n$ system. As an {\it ansatz}, we assume that a $T=0$ seniority-zero eigenstate of the Hamiltonian (\ref{1}) is a linear superposition of states which are products of four pairs (\ref{4}) arranged into $T=0$ quartets. These pairs must be symmetrically distributed among the quartets.
To express this more formally, we introduce the space of states
\begin{equation}
S^{(4)}=\Bigl\{
|1\rangle =[B^\dagger_1B^\dagger_2]^0[B^\dagger_3B^\dagger_4]^0|0\rangle ,
|2\rangle =[B^\dagger_1B^\dagger_3]^0[B^\dagger_2B^\dagger_4]^0|0\rangle ,
|3\rangle =[B^\dagger_1B^\dagger_4]^0[B^\dagger_2B^\dagger_3]^0|0\rangle
\Bigr\} ,
\label{15}
\end{equation}
where the distribution of the pairs within the quartets is such to leave invariant the space under the interchange of any two pairs.
Each eigenstate has therefore the form
\begin{equation}
|\Psi^{(4)}\rangle =d_1|1\rangle + d_2|2\rangle +d_3|3\rangle .
\label{16}
\end{equation}
One finds that
\begin{equation}
H|\Psi^{(4)}\rangle=(\sum^4_{\nu=1}E_\nu)|\Psi^{(4)}\rangle
+\sum^{3}_{s=1}\sum^4_{\nu=1}\biggl\{d_s-g\sum_i\frac{d_s}{2\epsilon_i-E_\nu}-g\sum_{\nu '\neq\nu}\frac{S_{\nu '\nu}(s)}{E_\nu '-E_\nu}\biggr\}|s_\nu\rangle
\label{18}
\end{equation}
where the state $|s_\nu\rangle$ is obtained from the state $|s\rangle$ of $S^{(4)}$ by replacing $B^\dagger_\nu$ with $P^\dagger$ (\ref{19}). An explicit version of this expression and the definition of the matrices $S_{\nu '\nu}(s)$ appropriate for this case can be found in the Appendix.

One deduces from Eq. (\ref{18}) that 
$|\Psi^{(4)}\rangle$ is an eigenstate of $H$ with eigenvalue  $E^{(4)}=\sum^4_{\nu=1}E_\nu$ if the twelve polynomials in curly brackets are zero. This defines a system of twelve coupled non-linear equations of the type
\begin{equation}
f(d_s,E_{\nu})=\frac{d_s}{g}-\sum_i\frac{d_s}{2\epsilon_i-E_{\nu}}-
\sum_{\nu '\neq\nu}\frac{S_{\nu '\nu}(s)}{E_{\nu '}-E_\nu}=0.
\label{7}
\end{equation}
The variables involved in these equations are the three amplitudes $d_s$ and the four pair energies $E_\nu$. Since the normalization of $|\Psi^{(4)}\rangle$ is unimportant, we can set one of the amplitudes, say $d_1$, equal to 1 and we are therefore left with the six unknowns $d_2,d_3,E_1,E_2,E_3,E_4$. The system to be solved in this case is therefore made of twelve equations in these six unknowns.
The solution of such an overdetermined system of equations cannot be guaranteed in principle. However, in the hypothesis that such a solution exists, this must be necessarily such that
\begin{equation}
\sum^{3}_{s=1}\sum^4_{\nu =1}f(d_s,E_{\nu})^2=0.
\label{21}
\end{equation}
It follows that such a solution is found if the minimum of the function
\begin{equation}
F=\sum^{3}_{s=1}\sum^4_{\nu =1}f(d_s,E_{\nu})^2
\label{22}
\end{equation} 
with respect to the variables $d_s,E_{\nu}$ is zero. This guarantees indeed that all the equations of the system have been exactly satisfied. The variables found in correspondence with such a minimum provide the solution of the system. 
This is the approach that has been followed to solve the system of equations generated by our formalism in all the cases that we have treated in this paper.

The formalism described so far for the cases of $2p-2n$ and $4p-4n$ systems can be extended to a generic $Np-Nn$ system.
We define in this case a space of states
\begin{equation}
{\it S}^{(N)}=\Biggl\{
|s\rangle =\prod^{N/2}_{q=1}[B^\dagger_{\nu (1,q,s)}B^\dagger_{\nu (2,q,s)}]^0|0\rangle
\Biggr\}_{s=1,2,\cdots ,N_s}
\label{5}
\end{equation}
where, as in the case of ${\it S}^{(4)}$(\ref{15}),
each of the $N_s$ states $|s\rangle$ of this space is a product of $N$ pairs (\ref{4})
arranged into $T=0$ quartets and where the distribution of the $N$ pairs among the quartets is chosen such to leave invariant the space under the interchange of any two pairs.  
As an {\it ansatz} for a $T=0$ seniority-zero eigenstate of the Hamiltonian (\ref{1}) for a 
$2N$-particle system we assume then the state
\begin{equation}
|\Psi^{(N)}\rangle =\sum^{N_s}_{s=1}d_s|s\rangle .
\label{6}
\end{equation}
with $|s\rangle$ belonging to ${\it S^{(N)}}$.
This state depends on two sets of variables: the $N$ pair energies $E_\nu$ of (\ref{4}) and the $N_s$ amplitudes $d_s$ of the expansion (\ref{6}). 
Choosing these variables such to satisfy the set of coupled non-linear equations still of the type (\ref{7}) for $d_s=(1,N_s)$ and $\nu =(1,N)$
guarantees the identity
\begin{equation}
H|\Psi^{(N)}\rangle =(\sum^N_{\nu =1}E_\nu )|\Psi^{(N)}\rangle ,
\label{8}
\end{equation}
namely $|\Psi^{(N)}\rangle$ is an eigenstate of $H$ and the associated eigenvalue is just the sum of the $N$ pair energies $E_\nu$. This occurs because, for a generic $Np-Nn$ system, it is
\begin{equation}
H|\Psi^{(N)}\rangle=(\sum^N_{\nu=1}E_\nu)|\Psi^{(N)}\rangle
+\sum^{N_s}_{s=1}\sum^N_{\nu=1}\biggl\{d_s-g\sum_i\frac{d_s}{2\epsilon_i-E_\nu}-g\sum_{\nu '\neq\nu}\frac{S_{\nu '\nu}(s)}{E_\nu '-E_\nu}\biggr\}|s_\nu\rangle .
\label{20}
\end{equation}
This expression represents the natural generalization of Eq. (\ref{18}) for a N-pair system.
For what concerns the matrices $S_{\nu '\nu}(s)$ to be used in a general case,
we remark that providing an analytical expression of these matrices requires the definition of  the $N_s$ states $|s\rangle$ of $S^{(N)}$ (\ref{5}) and, in correspondence with each of them, the definition of a matrix formulated in terms of the amplitudes $d_s$ of the expansion (\ref{6}). This has been done for $N=4$ but showing explicitly all these matrices for systems with $N>4$ pairs becomes hardly possible owing to the size of $N_s$ (see below). In the following we state a practical rule to construct these matrices for a
generic $Np-Nn$ system. In order to generate the element $S_{\nu '\nu}(s)$,
two cases have to be distinguished: (I), the quartet 
$[B^\dagger_{\nu '}B^\dagger_{\nu}]^0$ belongs to the
state $|s\rangle$ or, (II), it does not. In case (I), the element $S_{\nu '\nu}(s)$ is the sum of the amplitude $d_s$ plus the amplitudes of all those states of $S^{(N)}$ where the pairs $B^\dag_\nu$ and $B^\dag_{\nu'}$ belong to distinct quartets while the remaining quartets, if any, are the same as in $|s\rangle$. In case (II), instead, $S_{\nu '\nu}(s)$ includes only the sum of $d_s$ plus the amplitude $d_{s'}$ associated with the state $|s'\rangle$ which is obtained by interchanging the pairs $B^\dag_\nu$ and $B^\dag_{\nu'}$ in $|s\rangle$. This sum has to be taken with a reversed sign. The application of this rule generates, in particular, the matrices $S_{\nu '\nu}(s)$ for the $4p-4n$ case which are shown in the Appendix.

It is worthy noticing that the equations $(\ref{7})$ closely remind
the analogous equations introduced by Richardson in the case of like-particle pairing \cite{richa0,richa3,richa2} by exactly reducing to these if $d_s\equiv 1$ and $S_{\nu '\nu}(s)\equiv -2$. However, as compared with that case and as already discussed for the $4p-4n$ system, the explicit dependence on the amplitudes $d_s$ of the equations $(\ref{7})$ causes the total number of these equations to be larger than that
of the variables ($d_s$,$E_\nu$), the two numbers being  $N_{eq}=N_s\cdot N$ and $N_v=N_s +N-1$, respectively (where, in this definition of $N_v$, we have taken into account the irrelevancy of the normalization of $|\Psi^{(N)}\rangle$).
Only in the simple case of a $2p-2n$ system one has that $N_v=N_{eq}$.
As already anticipated, in order to solve this overdetermined system of $N_{eq}$ equations in $N_v$ unknowns, in all the cases that we have treated in this paper we have followed the approach illustrated for the $4p-4n$ system and therefore searched for the minimum of the function  
\begin{equation}
F=\sum^{N_s}_{s=1}\sum^N_{\nu =1}f(d_s,E_{\nu})^2
\label{23}
\end{equation} 
with respect to the variables $d_s,E_{\nu}$. In all cases the exact solution has been found.

\section{Numerical applications}
The systems that we have studied represent prototypes of axially deformed self-conjugate nuclei. They consist of $4\cal{N}$ nucleons, $2\cal{N}$ protons and $2\cal{N}$ neutrons, distributed over a set of $2\cal{N}$ equispaced, fourfold degenerate levels (i.e. we have considered only half-filled systems). To keep the model somehow close to some realistic calculations performed in the $sd$ shell \cite{samba}, we have assumed single-particle energies $\epsilon_i =-16+2(i-1)$, what corresponds to a spacing $\Delta\epsilon =2$. In the present calculations both these energies and the pairing strength $g$ will be expressed in arbitrary units.

Before discussing some numerical results, we have to remark that, similarly to what happens for the like-particle pairing \cite{richa0,richa3,richa2}, different eigenstates can be generated from equations (\ref{7}) in correspondence with different ``boundary conditions". For the ground state of a $4\cal{N}$-nucleon system, for instance, we require that, at $g=0$, the lowest $\cal{N}$ levels are fully occupied, each level by an uncorrelated $T=0$ quartet of the type $[P^\dag_\nu P^\dag_\nu]^0$, and the remaining $\cal{N}$ levels are empty. 
This is realized in practice by letting the ground state approach, in the limit $g\rightarrow 0$, a single  product of quartets, the two pair energies associated with each quartet pointing toward one of the values $2\epsilon_i$ $(i=1,2,\cdots ,\cal{N})$.
Excited states result from different ``initial" (i.e. $g\rightarrow 0$) configurations of the quartets. 
The total number of linearly independent 
configurations of this type provides the actual number of eigenstates which can be generated from equations (\ref{7}).
As we have also verified, this number coincides with the number of $T=0$ eigenvalues which are found by diagonalizing $H$ in the space of states $(\ref{3})$ therefore exhausting the total number of $T=0$ seniority-zero eigenstates of the model.
\begin{figure}[ht]
\begin{center}
\includegraphics[width=4.2in,height=4.9in,angle=-90]{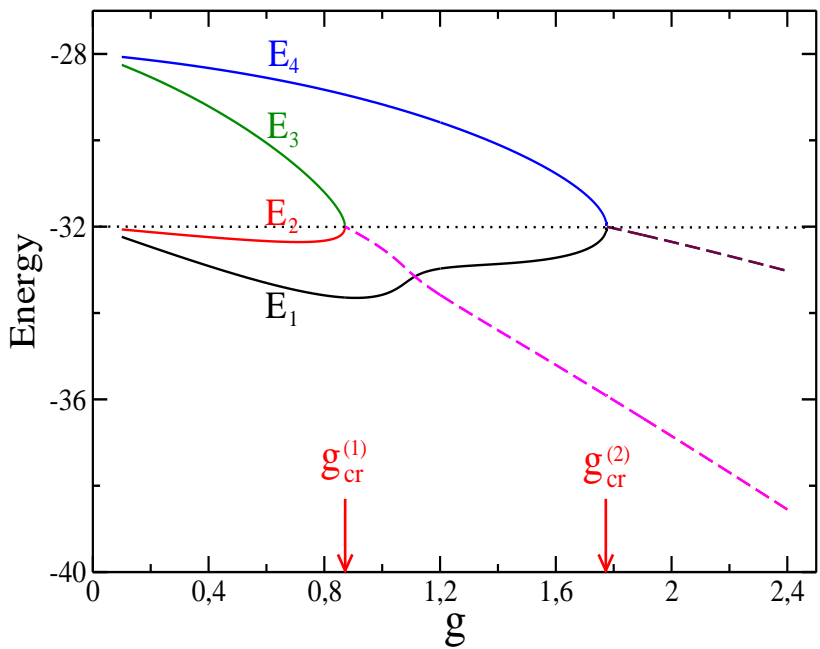}
\caption{(Color online) Pair energies relative to the ground state of a system with 4 protons and 4 neutrons as a function of the pairing strength $g$. Pair energies are real (solid line) up to the point where they come together (critical point). At this point they turn from real into complex-conjugate. The dashed line starting at the critical point denotes the common real part. All quantities are in arbitrary units.}
\label{f1}
\end{center}
\end{figure}
\begin{figure}[ht]
\begin{center}
\includegraphics[width=4.2in,height=4.9in,angle=-90]{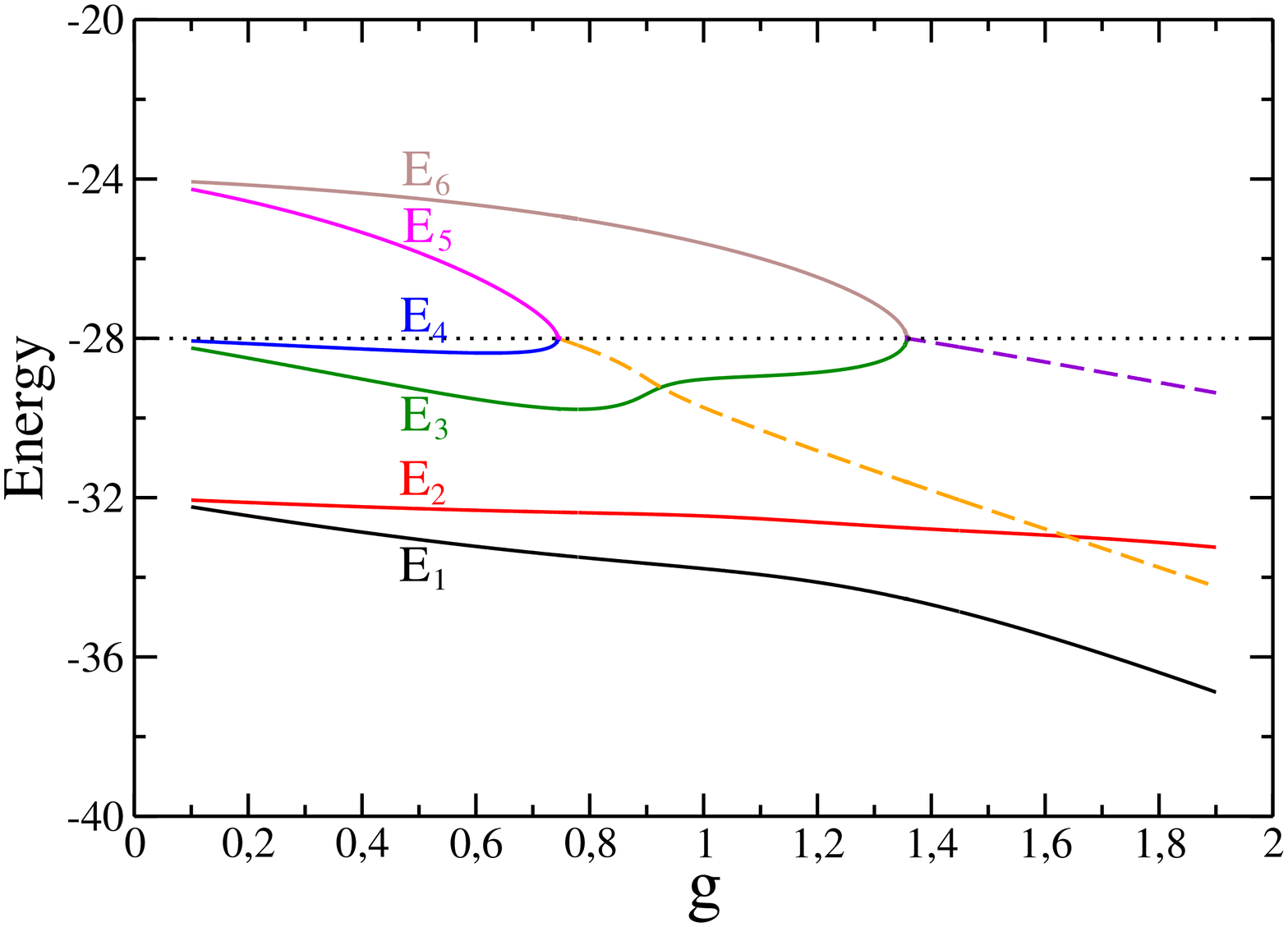}
\caption{(Color online) The same as in Fig. 1 for a system with 6 protons and 6 neutrons.}
\label{f2}
\end{center}
\end{figure}
\begin{figure}[ht]
\begin{center}
\includegraphics[width=4.2in,height=4.9in,angle=-90]{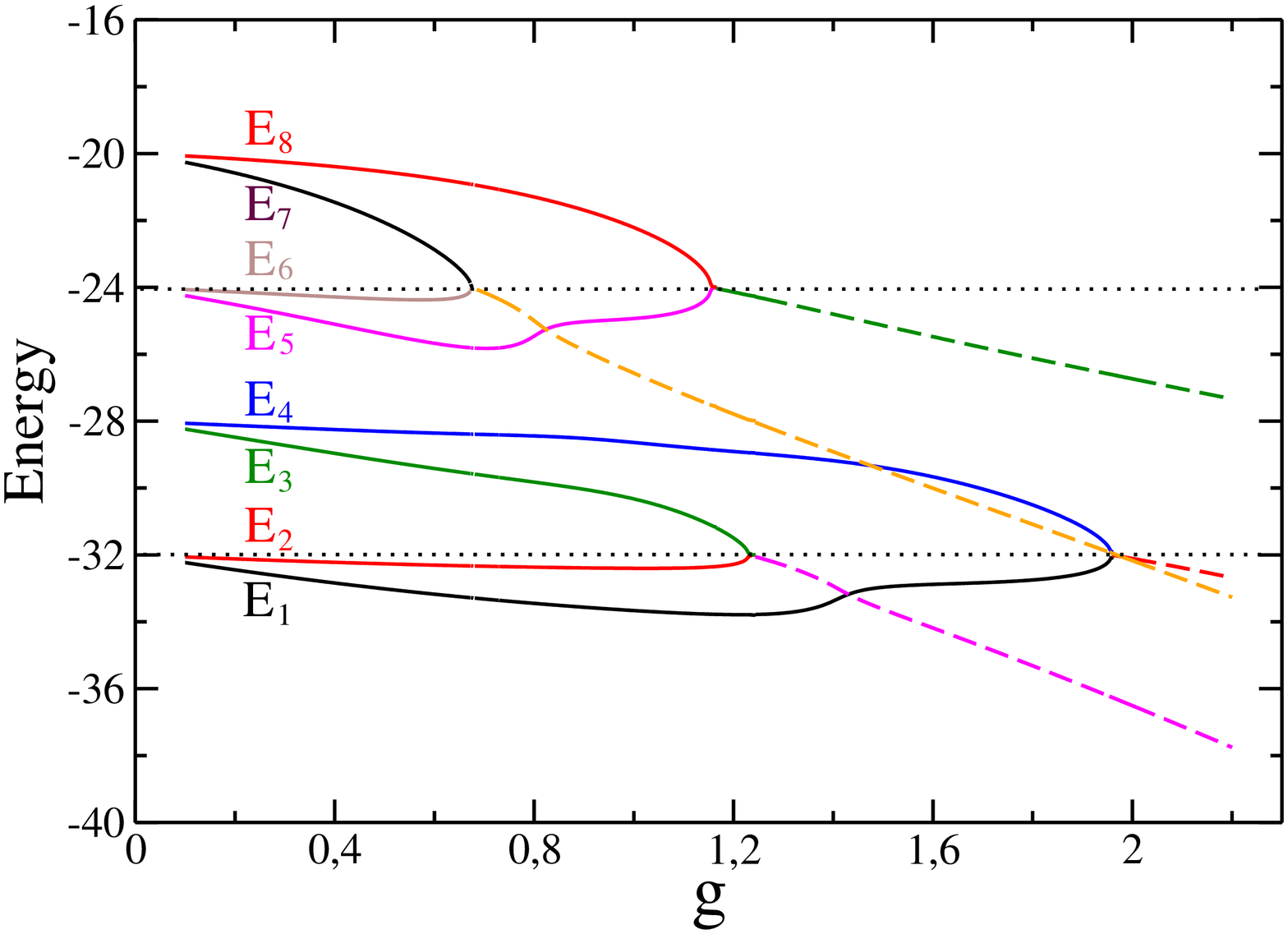}
\caption{(Color online) The same as in Fig. 1 for a system with 8 protons and 8 neutrons.}
\label{f3}
\end{center}
\end{figure}
\begin{figure}[ht]
\begin{center}
\includegraphics[width=4.2in,height=4.9in,angle=-90]{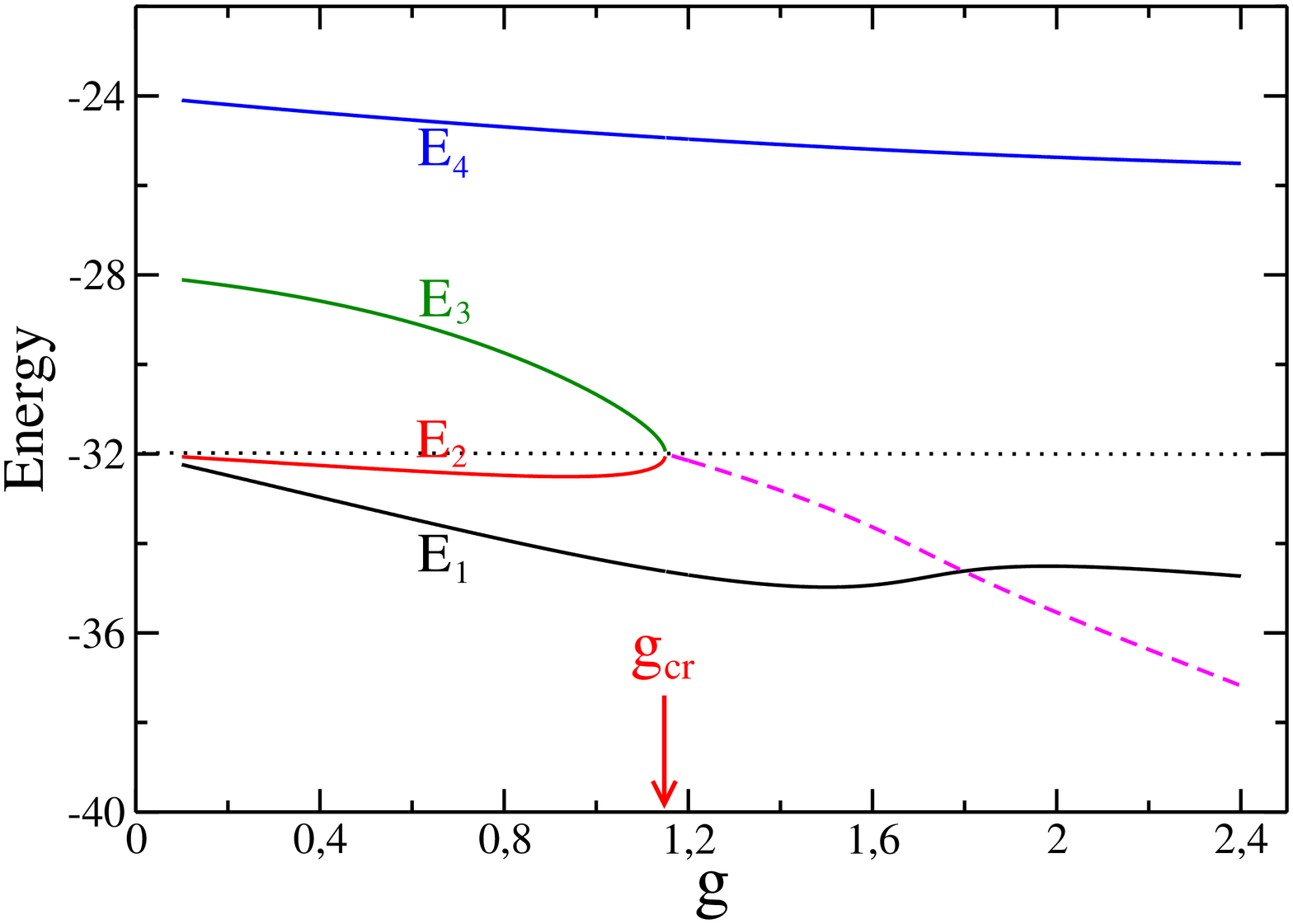}
\caption{(Color online) Pair energies relative to the first excited state of a system with 4 protons and 4 neutrons. Solid and dashed lines have the same meaning as in Fig. 1.}
\label{f4}
\end{center}
\end{figure}

In Fig. 1 we show the pair energies $E_\nu$ which characterize the ground state of a system of 4 protons and 4 neutrons (i.e. a two-quartet system). This case is simple enough to be discussed in some detail. As already seen in Sec. 2, the ground state is represented by the state $|\Psi^{(4)}\rangle$ (\ref{16}).
In this case, as well as in all the following applications of our formalism, when searching for the solutions of equations (\ref{7}) we have allowed both the amplitudes $d_s$ and the pair energies $E_\nu$ to be complex numbers.
Two critical points, $g^{(1)}_{cr}$ $(\approx 0.88)$ and $g^{(2)}_{cr}$ $(\approx 1.78)$,
are observed in Fig. 1. 
For $g<g^{(1)}_{cr}$,  all variables $d_s$ and $E_\nu$ turn out to be real. In the limit $g\rightarrow 0$, in particular, one finds that $d_1\rightarrow 1$ and
$d_2,d_3\rightarrow 0$ while the four pair energies are seen to converge
two-by-two toward the two lowest values $2\epsilon_i$. In this limit, then,
$|\Psi^{(4)}\rangle\rightarrow [P^\dag_1P^\dag_1]^0[P^\dag_2P^\dag_2]^0|0\rangle$.
For $g\rightarrow g^{(1)}_{cr}$, two pair energies belonging to pairs initially sitting on different 
single-particle levels are seen to approach the same value (coinciding with the smallest of the two initial values $2\epsilon_i$) and, exactly at $g=g^{(1)}_{cr}$, they turn from real into complex-conjugate quantities. The dashed line starting at this point in Fig. 1 illustrates their common real part after the critical point. This mechanism of entanglement between pairs is identical to that observed in the case of like-particle pairing \cite{richa0,richa3,richa2}. The two remaining pair energies suffer the same fate at $g^{(2)}_{cr}$. In the interval between the two critical strengths also the
amplitudes $d_1$ and $d_2$ become complex-conjugate while $d_3$ remains real. 
Finally, for
$g>g^{(2)}_{cr}$, the four pair energies evolve in a complex-conjugate form with all the amplitudes $d_s$ being real. We remark that, at each critical point, the system of equations (\ref{7}) undergoes a numerical instability. In Ref. \cite{richa0,richa3,richa2}, analogous instabilities were cured through some transformations of variables. In the present work, we 
have limited ourselves to approach these points from the left and from the right by proceeding in very small steps of $g$.

In Fig. 2 we show the pair energies characterizing the ground state of a system of 6 protons and 6 neutrons (i.e. a three-quartet system). 
The expansion (\ref{6}) counts $N_s=15$ components in this case .
Similarly to the two-quartet case, for $g\rightarrow 0$, the six pair energies converge two-by-two toward the lowest three values $2\epsilon_i$.  
The four uppermost pair energies in this limit are therefore seen to behave as
in Fig. 1 with increasing $g$. In the same range of the pairing strength, the two remaining pair energies remain, instead, always real and well distinct.

As a final example of a ground state,
Fig. 3 shows the pair energies relative to a system of 8 protons and 8 neutrons (i.e. a four-quartet system). 
The expansion (\ref{6}) counts $N_s=105$ components for this system.
The eight pair energies clearly split into two groups of four, each group behaving as in Fig. 1. The dependence of these pair energies on the pairing strength
is rather similar to the one observed for a system of 8 identical particles \cite{richa2}, the basic difference arising from the fact that in the present case two pairs can occupy each single-particle level. As a result, the pattern of the pair energies for the proton-neutron system appears as ``doubled" with respect to the like-particle case.

All the cases discussed so far have concerned the ground state of a $N=Z$ system.
As an example for an excited state, in Fig. 4 we show the pair energies which characterize the first excited state of a system with 4 protons and 4 neutrons. 
The wave function of this state is still represented by Eq. (\ref{16}).
At variance with the corresponding case of Fig. 1, only one critical point $g_{cr}$ ($\approx 1.15$) is observed. For $g<g_{cr}$, all the variables $d_s$ and $E_\nu$ are found to be real. In the limit $g\rightarrow 0$, in particular, one finds that
$|\Psi^{(4)}\rangle\rightarrow [P^\dag_1P^\dag_1]^0[P^\dag_2P^\dag_3]^0|0\rangle$. For $g\rightarrow g_{cr}$, two pair energies which belong to pairs initially sitting on the two lowest single-particle levels approach each other and, at $g=g_{cr}$, they become equal (also in this case this value coincides with the smallest of the two initial values $2\epsilon_i$) and turn from real into complex-conjugate quantities. The same transformation is undergone, at the same point, by the amplitudes $d_1$ and $d_2$ while $d_3$ remains always real. No further transformation is observed for $g>g_{cr}$.

\section{Conclusions}
In this paper we have derived the exact $T=0$ seniority-zero eigenstates of the isovector pairing Hamiltonian for even numbers of protons and neutrons distributed over a set of non-degenerate levels and interacting through a pairing force with constant strength. Various numerical applications have been provided which concern both ground and excited states. Two are the key features which have been clearly identified in these eigenstates: a), the presence of $T=1$ collective pairs acting as building blocks (these pairs having a similar form and behavior as the Richardson pairs  of the like-particle pairing)
and, b), the coupling of these $T=1$ pairs into $T=0$ quartets.
A similar grouping of protons and neutrons first into $T=1$ collective pairs and then into $T=0$ quartets had been assumed in previous approximate treatments of the isovector pairing Hamiltonian\cite{nicu,samba}. 
Moreover, the role of $T=0$ quartets had also emerged in an analysis of this Hamiltonian under the simplifying assumption  of degenerate single-particle levels (the so-called SO(5) model) \cite{dobes}.
However, the fact that the exact eigenstates of the isovector pairing Hamiltonian in the realistic case of non-degenerate single-particle levels
could be formulated in terms of $T=0$ quartets built with Richardson pairs 
had escaped the previous exact treatments of this Hamiltonian. As a major result, then,
the isovector pairing force emerges from this work as a key element to understand the formation of $\alpha$-like structures in $N=Z$ nuclei. 

\begin{acknowledgments}
This work was supported by a grant of Romanian Ministry of Research and Innovation, CNCS - UEFISCDI, project number
PN-III-P4-ID-PCE-2016-0481, within PNCDI III.
\end{acknowledgments}

\newpage
\section{appendix}
\begin{eqnarray}
H|\Psi^{(4)}\rangle&=&(E_1+E_2+E_3+E_4)|\Psi^{(4)}\rangle\nonumber\\
&&+\biggl\{d_1-g\sum_i\frac{d_1}{2\epsilon_i-E_1}-g\sum_{\nu\neq 1}\frac{S_{\nu 1}(1)}{E_\nu -E_1}\biggr\}
[P^\dagger B^\dagger_2 ]^0[B^\dagger_3B^\dagger_4]^0|0\rangle\nonumber\\
&&+\biggl\{d_1-g\sum_i\frac{d_1}{2\epsilon_i-E_2}-g\sum_{\nu\neq 2}\frac{S_{\nu 2}(1)}{E_\nu -E_2}\biggr\}
[P^\dagger B^\dagger_1 ]^0[B^\dagger_3B^\dagger_4]^0|0\rangle\nonumber\\
&&+\biggl\{d_1-g\sum_i\frac{d_1}{2\epsilon_i-E_3}-g\sum_{\nu\neq 3}\frac{S_{\nu 3}(1)}{E_\nu -E_3}\biggr\}
[P^\dagger B^\dagger_4 ]^0[B^\dagger_1B^\dagger_2]^0|0\rangle\nonumber\\
&&+\biggl\{d_1-g\sum_i\frac{d_1}{2\epsilon_i-E_4}-g\sum_{\nu\neq 4}\frac{S_{\nu 4}(1)}{E_\nu -E_4}\biggr\}
[P^\dagger B^\dagger_3 ]^0[B^\dagger_1B^\dagger_2]^0|0\rangle\nonumber\\
&&+\biggl\{d_2-g\sum_i\frac{d_2}{2\epsilon_i-E_1}-g\sum_{\nu\neq 1}\frac{S_{\nu 1}(2)}{E_\nu -E_1}\biggr\}
[P^\dagger B^\dagger_3 ]^0[B^\dagger_2B^\dagger_4]^0|0\rangle\nonumber\\
&&+\biggl\{d_2-g\sum_i\frac{d_2}{2\epsilon_i-E_2}-g\sum_{\nu\neq 2}\frac{S_{\nu 2}(2)}{E_\nu -E_2}\biggr\}
[P^\dagger B^\dagger_4 ]^0[B^\dagger_1B^\dagger_3]^0|0\rangle\nonumber\\
&&+\biggl\{d_2-g\sum_i\frac{d_2}{2\epsilon_i-E_3}-g\sum_{\nu\neq 3}\frac{S_{\nu 3}(2)}{E_\nu -E_3}\biggr\}
[P^\dagger B^\dagger_1 ]^0[B^\dagger_2B^\dagger_4]^0|0\rangle\nonumber\\
&&+\biggl\{d_2-g\sum_i\frac{d_2}{2\epsilon_i-E_4}-g\sum_{\nu\neq 4}\frac{S_{\nu 4}(2)}{E_\nu -E_4}\biggr\}
[P^\dagger B^\dagger_2 ]^0[B^\dagger_1B^\dagger_3]^0|0\rangle\nonumber\\
&&+\biggl\{d_3-g\sum_i\frac{d_3}{2\epsilon_i-E_1}-g\sum_{\nu\neq 1}\frac{S_{\nu 1}(3)}{E_\nu -E_1}\biggr\}
[P^\dagger B^\dagger_4 ]^0[B^\dagger_2B^\dagger_3]^0|0\rangle\nonumber\\
&&+\biggl\{d_3-g\sum_i\frac{d_3}{2\epsilon_i-E_2}-g\sum_{\nu\neq 2}\frac{S_{\nu 2}(3)}{E_\nu -E_2}\biggr\}
[P^\dagger B^\dagger_3 ]^0[B^\dagger_1B^\dagger_4]^0|0\rangle\nonumber\\
&&+\biggl\{d_3-g\sum_i\frac{d_3}{2\epsilon_i-E_3}-g\sum_{\nu\neq 3}\frac{S_{\nu 3}(3)}{E_\nu -E_3}\biggr\}
[P^\dagger B^\dagger_2 ]^0[B^\dagger_1B^\dagger_4]^0|0\rangle\nonumber\\
&&+\biggl\{d_3-g\sum_i\frac{d_3}{2\epsilon_i-E_4}-g\sum_{\nu\neq 4}\frac{S_{\nu 4}(3)}{E_\nu -E_4}\biggr\}
[P^\dagger B^\dagger_1 ]^0[B^\dagger_2B^\dagger_3]^0|0\rangle .\nonumber
%
\end{eqnarray}
The matrices $S_{\nu '\nu} (s)$  entering this expression are defined in the following Tables I-III. 
\begin{table}[h]
\caption{The matrix $S_{\nu '\nu} (1)$.}
\begin{center}
\begin{tabular}{|cccc|}
    -     &      ($d_1+d_2+d_3$)      &     ($-d_1-d_3$)       &     ($-d_1-d_2$)   \\
($d_1+d_2+d_3$)   &      -       &      ($-d_1-d_2$)   &  ($-d_1-d_3$)  \\    
($-d_1-d_3$)  &    ($-d_1-d_2$)   &    -    &    ($d_1+d_2+d_3$) \\
($-d_1-d_2$) &  ($-d_1-d_3$)  &  ($d_1+d_2+d_3$)  &  -  \\
\end{tabular}
\end{center}
\end{table}
\begin{table}[h]
\caption{The matrix $S_{\nu '\nu} (2)$.}
\begin{center}
\begin{tabular}{|cccc|}
    -     &      ($-d_2-d_3$)      &     ($d_1+d_2+d_3$)       &     ($-d_1-d_2$)   \\
($-d_2-d_3$)   &      -       &      ($-d_1-d_2$)   &  ($d_1+d_2+d_3$)  \\    
($d_1+d_2+d_3$)  &    ($-d_1-d_2$)   &    -    &    ($-d_2-d_3$) \\
($-d_1-d_2$) &  ($d_1+d_2+d_3$)  &  ($-d_2-d_3$)  &  -  \\
\end{tabular}
\end{center}
\end{table}
\begin{table}[h]
\caption{The matrix $S_{\nu '\nu} (3)$.}
\begin{center}
\begin{tabular}{|cccc|}
    -     &      ($-d_2-d_3$)      &     ($-d_1-d_3$)       &     ($d_1+d_2+d_3$)   \\
($-d_2-d_3$)   &      -       &      ($d_1+d_2+d_3$)   &  ($-d_1-d_3$)  \\    
($-d_1-d_3$)  &    ($d_1+d_2+d_3$)   &    -    &    ($-d_2-d_3$) \\
($d_1+d_2+d_3$) &  ($-d_1-d_3$)  &  ($-d_2-d_3$)  &  -  \\
\end{tabular}
\end{center}
\end{table}

\end{document}